\documentclass[aip,jap,reprint, amsmath,amssymb]{revtex4-1}

\usepackage{graphicx}
\usepackage{dcolumn}
\usepackage{bm}
\usepackage{gensymb}

\begin{document}


\title{Effect of device design on charge offset drift in Si/SiO$_2$ single electron devices}


\author{Binhui Hu}
\email[To whom correspondence should be addressed. Electronic mail:
] {hubh@umd.edu, stew@nist.gov}
 \affiliation{Joint Quantum Institute, University of Maryland, College Park, Maryland 20742, USA}
 \author{Erick D. Ochoa}
 \author{Daniel Sanchez}
 \author{Justin K. Perron}
 \affiliation{California State University-San Marcos, San Marcos, CA 92069, USA}
\author{Neil M. Zimmerman}
\author{M. D. Stewart, Jr.$^*$}
\affiliation{National Institute of Standards and Technology, Gaithersburg, MD, 20899, USA}


\date{\today}

\begin{abstract}
We have measured the low-frequency time instability known as charge offset drift of Si/SiO$_2$ single electron devices (SEDs) with and without an overall poly-Si top gate. We find that SEDs with a poly-Si top gate have significantly less charge offset drift, exhibiting fewer isolated jumps and a factor of two reduction in fluctuations about a stable mean value. The observed reduction can be accounted for by the electrostatic reduction in the mutual capacitance $C_m$ between defects and the quantum dot, and increase in the total defect capacitance $C_d$ due to the top gate. These results depart from the accepted understanding that the level of charge offset drift in SEDs is determined by the intrinsic material properties, forcing consideration of the device design as well. We expect these results to be of importance in developing SEDs for applications from quantum information to metrology or wherever charge noise or integrability of devices is a challenge.
\end{abstract}


\maketitle

\section{INTRODUCTION}
Single electron devices (SEDs) have many important applications due to their ability to localize and manipulate individual electrons' degrees of freedom. SEDs have been proposed as current standards in electrical metrology,\cite{Keller96, Wright09, Maisi09, Fletcher12, Giblin2012, Pekola13, Yamahata2014, Nakamura15,Yamahata15} and as memory and logical devices in integrated circuits.\cite{chen95, Likharev99, Takahashi02, Takahashi04, Maeda12} They have also been studied as qubits, when there are only a few electrons on the quantum dot.\cite{Loss98, Kane1998, Wiel02, Hanson07, Zwanenburg13} Ultimately, these applications must integrate many SEDs together, which, in turn, requires the operating point of each SED to remain stable in time. However, a long-standing, low-frequency time instability known as charge offset drift $Q_0(t)$ present in real devices remains a challenge to realizing the full potential of SEDs.\cite{Stew16}

Up to now, charge offset drift was understood to be a consequence of intrinsic material properties.\cite{Neil08JAP, Stew16} This material system explanation is based on the experimental fact that SEDs made in two material systems have very different $Q_0(t)$: Al/AlO$_x$-based SEDs exhibit a large change in charge offset ($\Delta Q_0>1\,e$), while in mesa-etched Si/SiO$_2$-based silicon-on-insulator (SOI) devices the change of charge offset is small ($\Delta Q_0<0.01\,e$). Here $e$ is the electron charge and the value of $Q_0(t)$ is normalized to the period of the Coulomb oscillations.\cite{Neil08JAP} This experimental fact is interpreted microscopically as a distinct difference in the level of interaction between two-level system (TLS) defects present in the amorphous insulators, AlO$_x$ and SiO$_2$. Specifically, TLS defects are not stable over time or gate voltage sweeps in Al/AlO$_x$ devices, while they are stable in Si/SiO$_2$ devices.\cite{Neil08JAP, Stew16} Unsuccessful attempts to reduce $\Delta Q_0$ in Al/AlO$_x$-based SEDs with different device geometries and structures lend additional weight to this materials-only explanation \cite{Huber01} which has had significant influence over the direction of SED research by emphasizing the expected performance edge implied for Si devices.\cite{Stew16} However, noise measurements have only been done in a very limited set of Si/SiO$_2$ based SED device structures: mainly mesa-etched SOI devices\cite{Neil2001, Neil2007} and Si SEDs with metal gates.\cite{Neil2014, Blake16}

Workers at Sandia National Labs and the University of Sherbrook have pioneered fabrication of Si/SiO$_2$-based SEDs with a single layer of doped polysilicon gates.\cite{Tracy13,Rudolph17} Those devices present an opportunity to assess the robustness of the above explanation for $Q_0(t)$ because they do not have a blanket gate to function as an SED, and their $Q_0(t)$ behavior differs from the mesa-etched SOI devices while not altering the Si/SiO$_2$ material system. \cite{Rudolph17} In this manuscript, we present data showing that Si/SiO$_2$ single gate layer devices show significantly larger $\Delta Q_0$ than their SOI counterparts and, moreover, that $\Delta Q_0$ can be reduced with the addition of another gate covering the fine area of the device. This design change mimics the gate stack of the very stable SOI devices mentioned above. Our results show that $Q_0(t)$ is not entirely determined by the material system as previously thought.

\section{SAMPLES AND EXPERIMENTAL DETAILS}

 \begin{figure}
 \includegraphics[width=8.5cm] {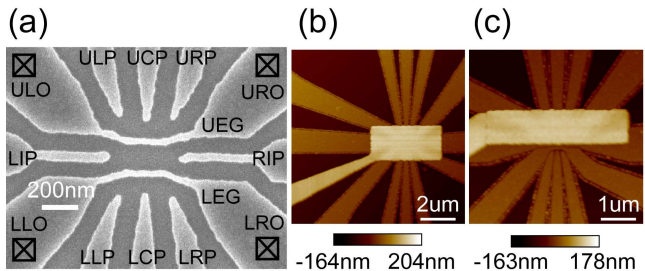}%
 \caption{\label{fig:device} The different types of devices employed in this study. (a) SEM image of a single gate layer device, showing poly-Si gates on gate oxide, referred to as ``bare''. Two individual single electron devices (SEDs) can be formed underneath positively biased enhancement gates UEG and LEG, while other plunger gates are used to define the quantum dot. (b) AFM image of an SED device with a full poly-Si top gate, referred to as ``TG''. (c) AFM image of an SED device with a half poly-Si top gate. The upper SED is covered by the poly-Si top gate (``TG''), and the lower one is not and is referred to as ``oxide''. Lower gates in these devices have the same geometry.}
 \end{figure}

We have fabricated and measured $Q_0(t)$ on three different types of devices shown in Fig.\ \ref{fig:device}. A single layer of poly-Si gates were patterned on top of a 37\,nm thick SiO$_2$ gate oxide. The gate arrangement shown in Fig.\ \ref{fig:device}(a), similar to Ref.~\onlinecite{Tracy13, Rudolph17}, produces two individual SEDs: one in the upper half of the image with the dot underneath the enhancement gate labeled UEG and one in the lower half of the image with the dot underneath the enhancement gate labeled LEG. On another wafer, after fabricating single layer devices, a 20\,nm thick isolation oxide was grown on the poly-Si gates, and then a full poly-Si top gate (Fig.\ \ref{fig:device}(b)) or a half poly-Si top gate (Fig.\ \ref{fig:device}(c)) was patterned on them. Between the two wafers we have three different types of SEDs: SEDs without isolation oxide or a top gate but with native oxide (Fig.\ \ref{fig:device}(a)), SEDs with isolation oxide and a top gate (Fig.\ \ref{fig:device}(b) and the upper half of Fig.\ \ref{fig:device}(c)), and SEDs with an isolation oxide but no top gate (the lower half of Fig.\ \ref{fig:device}(c)). We will refer to these devices as ``bare'', ``TG'', and ``oxide'', respectively. In total, we have measured two ``bare" devices, three ``TG" devices, and one ``oxide" device (Table~\ref{table:1}). This total includes a single die as depicted in Fig.\ \ref{fig:device}(c), where different types of SEDs lie within 200\,nm of each other.

All devices discussed here were fabricated at NIST on 150\,mm boron-doped silicon $<$100$>$ wafers with a resistivity of 5-10\,$\Omega\cdot$cm. The main fabrication process is as follows. The source/drain contacts (ULO, URO, LLO, LRO in Fig.\ \ref{fig:device}(a)) are formed by phosphorus ion implantation. Then, a 125\,nm field oxide is grown in a wet oxidation furnace at 900\,$\degree$C, and etched away in the device window (a 175\,$\mu$m square) using BOE.\cite{Rossi2015} Subsequently, a high-quality 37\,nm gate oxide is grown in a dry oxidation furnace with trichloroethane (TLC) at 950\,$\degree$C, which is immediately followed by deposition of a 75\,nm in-situ doped N$^+$ poly-Si layer at 625\,$\degree$C. The poly-Si gates are patterned by e-beam lithography using XR-1541 negative tone resist and a Cl$_2$-based dry etch. For ``bare'' devices, the next step is aluminum metallization to form contacts with a 425\,$\degree$C forming gas anneal as the last step. For ``oxide'' or ``TG'' devices, the first layer of gate lithography and etching is followed by growth of a 20\,nm isolation oxide on the poly-Si gates in a dry oxidation furnace at 850\,$\degree$C, and a second deposition of in-situ doped N$^+$ poly-Si. Then the top gate layer is defined using the same e-beam lithography process as for the lower gates. The final step is aluminum metallization and the 425\,$\degree$C forming gas anneal.

To measure the charge offset drift, a positive gate voltage is applied to the enhancement gate UEG (LEG) to accumulate electrons at the Si/SiO$_2$ interface, while other plunger gates are biased to define the quantum dot. The SED is tuned so that Coulomb blockade oscillations can be observed while sweeping either the enhancement gate or one of the plunger gates. We then repeatedly measure the same Coulomb blockade oscillation curve approximately every 15 minutes to track the changes in the local charge environment of the dot. $Q_0(t)$ is extracted from each trace using two different methods. At large source-drain bias (about half of the charging energy), the source-drain current $I_d$ oscillates sinusoidally when sweeping gate voltage $V_P$, and each trace is fit to a sinusoidal function: $I_d (V_P)=A_0+A\sin[2\pi(V_P/\Delta V_P +Q_0 (t)/e)]+BV_P$, where $A_0$ is a current offset, $A$ is the amplitude of the oscillations, $\Delta V_P$ is the period of the oscillations, and $B$ is used to account for any slope in the sinusoidal curve. If the trace is not sinusoidal, a Gaussian function is used to find the peak location $V_{Peak}(t)$, and $Q_0(t) =-e(V_{Peak}(t)-V_{Peak}(t=0))/\Delta V_P$, where $\Delta V_P$ is the average voltage difference between the peak of interest and the two neighboring peaks. We have used both methods to analyze sinusoidal traces, and found that the results are consistent within 10\%, predominantly due to the uncertainty associated with $\Delta V_P$. Measurements performed at NIST were taken at about 2.5\,K in a closed-cycle cryostat. To exclude the possibility that the experimental setup contributes to the measured drift, NIST measurements were performed with two different sets of electronics (discussed in Sect.\ \ref{sec:results}). Additional devices were also measured at CSUSM in another closed-cycle cryostat at about 2.5\,K. The results (see Table~\ref{table:1}) from each set of measurements are qualitatively and quantitatively similar to those presented for the same device type.

The measured quantum dots are not necessarily intentional quantum dots. We select the bias conditions and the data fitting range so that the device operates as a stable single quantum dot device, confirmed by two-dimensional gate voltage sweeps and Coulomb diamonds.\cite{binhui09} Each dot measured had a charging energy of about 5\,meV. As these quantum dots are used as local charge sensors of the environment, the detailed mechanism for the formation of the quantum dot is not expected to affect our conclusions.

\section{RESULTS} \label{sec:results}

\begin{figure}
 \includegraphics[width=8.5cm]{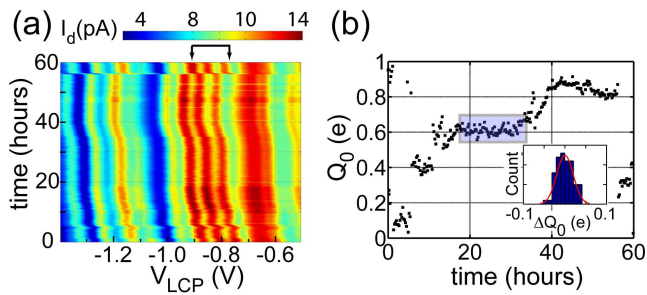}%
 \caption{\label{fig:nogate} $Q_0(t)$ for a ``bare'' device (4.7-41L). (a) Coulomb blockade oscillations taken with $V_{LEG}=3\,$V, $V_{LRP}=-1.8\,$V, $V_{RIP}=-2.48\,$V, all other gates at 0\,V, DC bias $V_{d-DC}=-2.5\,$mV and AC bias $V_{d-AC}=0.5\,$mV, while sweeping $V_{LCP}$ at T=2.5\,K using the AC measurement system (see text). (b) Charge offset drift $Q_0(t)$ vs time, extracted from the Coulomb blockade oscillations using a sinusoidal function in the range indicated by arrows as shown in (a). The inset shows a histogram of the deviation of $Q_0(t)$ from the mean value corresponding to the shaded area in the main panel. The histogram has a standard deviation $0.020\,e$.}
 \end{figure}

Figure \ref{fig:nogate} shows a typical result for a ``bare'' (device 4.7-41L) using standard AC lock-in amplifier techniques. The charge offset drift $Q_0(t)$ has three distinct features, which were also observed in devices of the same design fabricated at Sandia National Labs.\cite{Rudolph17} First, over the course of the first two days, $Q_0(t)$ shows an evolution from rapid drift toward slower drift while winding $Q_0(t)$ through several $e$. This phenomenon has been previously referred to as transient relaxation.\cite{Neil08JAP} (It should be noted that these single layer devices often become more stable at longer times.\cite{Rudolph17}) Second, the data show isolated discrete jumps or drifts, which are not stationary. Third, the device shows some stable periods where $Q_0(t)$ takes on a value within a stationary band. One such period is indicated by a shaded area in Fig.\ \ref{fig:nogate}(b). We characterize local fluctuations about a stable mean with the standard deviation $\sigma$. This metric, while useful in quantifying the differences between devices, does not capture discrete jumps or long-term drift; all three metrics affect device integrability.

\begin{figure} [h]
 \includegraphics[width=8.5cm]{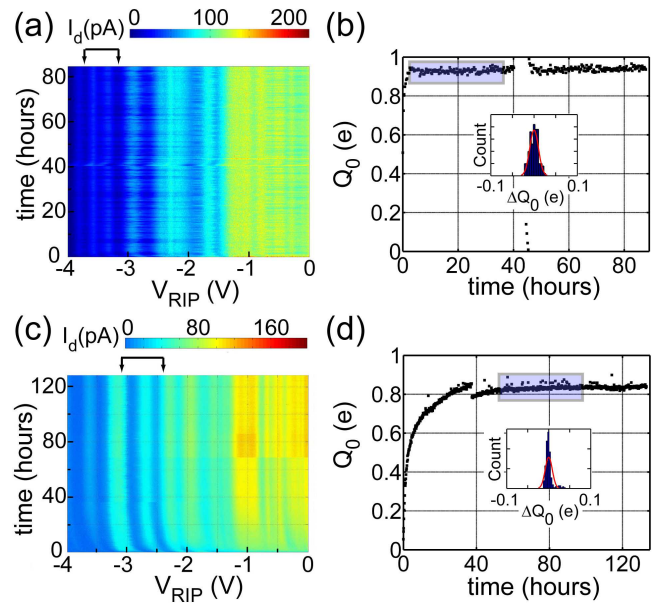}%
 \caption{\label{fig:gate} $Q_0(t)$ for a ``TG'' device (5.4-25U) using the DC measurement system and the AC measurement system at NIST. Data taken (a) at $V_{UEG}=1.42\,$V, $V_{TG}=0.2\,$V, $V_{LIP}=V_{ULP}=V_{UCP}= V_{URP}=0\,$V, $V_{d-DC}=3\,$mV using the DC measurement system and (c) at $V_{UEG}=1.4\,$V, $V_{TG}=0.2\,$V, $V_{LIP}=V_{UCP}=0\,$V, $V_{ULP}=V_{URP}=-0.9\,$V with $V_{d-DC}=2\,$mV and $V_{d-AC}=0.5\,$mV using the AC measurement system. Using sinusoidal functions in the ranges indicated by arrows, $Q_0(t)$ is extracted and shown in (b) and (d) respectively. The shaded areas show stable periods, and the insets show the histogram of the fluctuation $\Delta Q_0$  with standard deviations of $0.010\,e$ (DC) and $0.008\,e$ (AC) respectively.}
 \end{figure}

Figure \ref{fig:gate} shows $Q_0(t)$ behavior measured in a ``TG'' device. There are four main differences from the data depicted in Fig. \ref{fig:nogate}. First, while transient relaxation is still observed in the device with the top gate, $Q_0(t)$ winds through less than $1\,e$ variation before becoming stable and, in the first cooldown data (Fig \ref{fig:gate}(b)), reaches a stable value in just a few hours. Second, after the transient evolution, $Q_0(t)$ essentially remains stable for the duration of the measurement in the device with the top gate. Third, $Q_0(t)$ in top-gated devices shows fewer and smaller discrete jumps. Fourth, the local fluctuations about the stable mean value are reduced by more than a factor of two. All the devices of this type which we have measured show this behavior.

\begin{figure} [h]
 \includegraphics[width=8.5cm]{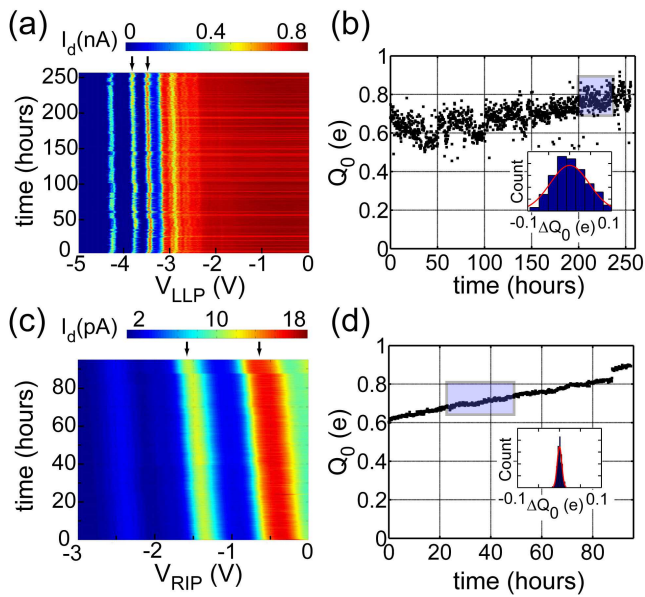}%
 \caption{\label{fig:halfgate} Comparison of $Q_0(t)$ on neighboring ``oxide'' and ``TG'' devices. (a) Coulomb blockade oscillations for ``oxide'' SED 5.4-11L taken at $V_{LEG}=5$V, $V_{LIP}=-2.8$V, all other gates at 0V, $V_{d-DC}=2$ mV using the DC measurement system, and (c) for ``TG'' SED 5.4-11U at $V_{UEG}=4$V, $V_{TG}=0.63$V, all other gates at 0V, $V_{d-DC}=3$ mV and $V_{d-AC}=0.5$ mV using the AC measurement system. (b) and (d) show the extracted $Q_0(t)$ using Gaussian functions. Arrows indicate the three peaks used in the fitting. The shaded areas show stable periods, and the insets show the histogram of the fluctuation $\Delta Q_0$ after subtracting a linear-fit line with standard deviations of $0.046\,e$ (5.4-11L) and $0.005\,e$ (5.4-11U).}
 \end{figure}

Finally, in an effort to investigate the role of the isolation oxide, we also measured neighboring ``oxide'' and ``TG'' SEDs. This also enables a comparison of devices within 200\,nm of each other in the same cooldown. Figure \ref{fig:halfgate} shows the measurement results. Interestingly, transient relaxation is absent in both devices while a systematic (approximately linear) drift is observed instead indicating a non-stationary process. The origin of the linear drift is not clear and requires further investigation. Notwithstanding the linear drift, we again find a reduction in the frequency and amplitude of discrete jumps as well as fluctuations in the ``TG'' device. To facilitate a comparison of the fluctuations with the previous data we fit the data to a line and remove this dependence before plotting the histograms shown in the insets. The $\sigma$ obtained this way shown in the ``oxide'' SED (device 5.4-11L) is about 9 times larger than that in the ``TG'' SED (device 5.4-11U). When assessed by this metric, the performance of the ``oxide'' device is the worst of those presented here even when compared to that of the ``bare'' device (see Fig. \ref{fig:nogate}). This may be due to the dry etching process used to remove the top gate, leaving behind some charge defects in the oxide.  When assessed by the long-term drift, the ``oxide'' device only shows small monotonic drift with few discrete jumps which is better than the ``bare'' device. In terms of both metrics, the ``TG'' device has the best performance.

A natural question when making these measurements is whether or not the measured drift is intrinsic to the device or from some extrinsic source in the measurement setup, especially since these measurements extend over days. To assess this question, measurements at NIST were performed with two separate sets of electronics and additional measurements were performed at CSUSM in a separate cryostat with a third set of electronics. The electronic systems at NIST are a DC measurement system using an Agilent 4156C precision semiconductor parameter analyzer, and an AC measurement system using standard lock-in amplifier techniques at 17\,Hz. The DC measurement electronics return all electrodes to zero voltage between measurements of the Coulomb blockade curve while the AC electronics keeps a steady voltage on each electrode while returning the swept electrode to the beginning of the sweep between measurements. The measurement electronics at CSUSM keep every electrode at the voltage corresponding to the end of the sweep between measurements. A comparison of the NIST measurement electronics is made in Fig. \ref{fig:gate} for an SED device with a top gate (device 5.4-25U). The DC system data are shown in Fig.\ \ref{fig:gate}(a)(b) and the AC system data are shown in Fig.\ \ref{fig:gate}(c)(d). Although the DC data are noisier than the AC data, they are qualitatively consistent with each other, and the standard deviation in the shaded areas is only different by 22$\%$. The longer transient relaxation time observed in the AC data is likely due to the intervening thermal cycle, which necessitated different applied voltages.

\section{DISCUSSION}

\begin{table*}[t]
\begin{center}
    \begin{tabular}{ | l | l | l | l | l | l |}
    \hline
\multicolumn{6}{|c|}{No top gate}\\
    \hline
Device &Type &Isolation Oxide&Measurement&Fitting &$\sigma$ of $Q_0(t) (e)$ \\
\hline
4.7-41L&``bare"&No&NIST AC&Sine&$0.020\pm0.004$\\
\hline
4.7-33U&``bare"&No&CSUSM DC&Gaussian&$0.017\pm0.003$ \\
\hline
5.4-11L&``oxide"&Yes&NIST DC&Gaussian&$0.046e\pm0.005$\\
\hline
\end{tabular}
\end{center}

\begin{center}
    \begin{tabular}{ | l | l | l | l |}
    \hline
\multicolumn{4}{|c|}{With top gate ``TG"}\\
    \hline
Device &Measurement& Fitting&$\sigma$ of $Q_0(t) (e)$\\
\hline
5.4-23U&NIST DC&Sine&$0.007\pm0.001$\\
\hline
5.4-25U&NIST DC& Sine&$0.010e\pm0.002$\\
\hline
5.4-25U&NIST AC& Sine&$0.008e\pm0.001$\\
\hline
5.4-11U&NIST AC&Gaussian&$0.005e\pm0.001$\\
\hline
\end{tabular}
\end{center}
\caption{Summary of measured charge offset drift $Q_0(t)$. Devices with a top gate exhibit about a factor of two lower $\sigma$ than those devices without a top gate. Uncertainty corresponds to a $95\%$ confidence interval and is estimated based on a $\chi^2$ distribution of $N-1$ degrees of freedom where $N$ corresponds to the number of data points.}
\label{table:1}
\end{table*}

The data are summarized in Table \ref{table:1} (additional data is shown in the Supplementary Material S3). In the context of the material-only explanation for $Q_0(t)$, all the devices listed should show similar drift; however, the data show that for Si/SiO$_2$-based SEDs, $Q_0(t)$ is influenced by factors other than the material system. We can characterize our empirical conclusions as follows: i) Long-term drift: ``bare'' devices (Fig.\ \ref{fig:nogate}) show random large-amplitude drift over the course of hours or days, similar to the Al/AlO$_x$ system,\cite{Neil08JAP, Huber01} ``oxide'' (Fig.\ \ref{fig:halfgate}(b)) and ``TG'' (Figures\ \ref{fig:gate} and \ref{fig:halfgate}(d)) devices show no drift or monotonic, predictable drift with fewer and smaller discrete jumps, similar to previous results in SOI devices;\cite{Neil2007} ii) Local fluctuations, i.e. $\sigma$: ``bare'' and ``oxide'' devices show substantially larger $\sigma$ than ``TG'' devices. Below we list possible mechanisms driving the difference in behavior shown in this manuscript.

There are clear electrical differences between the two types of devices. The top gate allows for the application of an electric field across the gate oxide in addition to the field resulting from the work function difference between the gate and the silicon. These fields could freeze defects out. The gate also acts as a ground plane which increases the total capacitance $C_d$ of charge defects in the SED, and at the same time decreases the mutual capacitance $C_m$ of defects coupled to the quantum dot. Considering a simplified case where one effective defect is coupled to the quantum dot, $Q_0(t) \approx (C_m/C_d)\Delta Q_d(t)$ (see the Supplementary Material S1), where $\Delta Q_d(t)$ is the variation of the defect charge. Both increasing the total capacitance of charge defects and decreasing the mutual capacitance to the quantum dot can reduce the charge offset drift $Q_0(t)$. To estimate the magnitude of the change in $C_m$ and $C_d$ (see the Supplementary Material S2), we have used FastCap to simulate ``bare'' and ``TG'' devices.\cite{Nabors91} For charge defects located 90\,nm away from the dot laterally and at the midpoint of the gate oxide thickness, the top gate reduces $C_m$ by 55\% and increases $C_d$ by 16\%. This reduces $C_m/C_d$ by about a factor of two which gives the same order of magnitude as the observed reduction in $\sigma$. For defects located nearer to the Si/SiO$_2$ interface, the reduction in $C_m/C_d$ is more muted, while it is more pronounced for defects near the SiO$_2$/gate interface. Similarly, for defects located nearer the quantum dot or other gates, the reduction in $C_m/C_d$ due to the top gate is more muted. The detailed spatial distribution of defects will impact the size of the change in $\sigma$, however, these calculations show that the observed reduction of $\sigma$ in $Q_0(t)$ can be accounted for by the electrostatic reduction in $C_m$ and increase in $C_d$, with the change in $C_m$ being dominant. Moreover, it is plausible that the lack of an exposed gate oxide surface for SOI devices and ``TG'' devices as compared to ``bare'' or ``oxide'' single-layer devices can effectively reduce the charge offset drift by reducing the number of defects near the gate oxide surface, where moisture or other ion defects may adsorb to the surface.

Additional electrical differences include the top poly-Si gate acting as a Faraday cage, which can shield the dot in the device from external electrical disturbance. The presence of the gate also necessarily implies additional strain in the device. Finally, though we have worked to minimize them, the gate cannot be introduced without some fabrication differences. These include second layer e-beam exposure, exposure to the dry etch process, and isolation oxide growth.

Whatever the reason for the reduction in $Q_0(t)$, these data indicate the previous understanding of charge offset drift as a material property is incomplete. We hasten to add, however, that the material stacks are still an important factor. In particular, as noted above, it appears that $Q_0(t)$ in ``bare'' devices (with a low-quality native oxide on the gates) have larger random drift and discrete jumps, similar to previous Al/AlO$_x$ results,\cite{Neil08JAP,Huber01} whereas the devices with deliberate high-quality isolation oxide show monotonic, predictable drift with few or no discrete jumps similar to previous SOI results.\cite{Neil2007} This may indicate that the native oxide, which can also pick up moisture from the surrounding air, has the same type of interacting defects that we have previously discussed.\cite{Neil08JAP} In addition to the intrinsic TLS defects, the long-term drift in the native oxide could also be due to such mechanisms as the movement of dissolved hydroxyl ions from adsorbed moisture. Further experiments are necessary to confirm this suggestion.

Earlier work on Al/AlO$_x$/Al based SEDs with a nano-Faraday cage did not show any improvement in the charge offset drift.\cite{Huber01} In fact, enclosing the device in additional AlO$_x$/Al increased $\Delta Q_0$ by approximately a factor of two, in striking contrast to the results presented here. The likely reason is that unlike the AlO$_x$/Al stack, the SiO$_2$/poly-Si top gate does not introduce a significant number of new unstable charge defects, so that the benefit of adding the gate as outlined above (which should not otherwise differ between the two material systems) outweighs the negative effect from additional charge defects.

\section{CONCLUSIONS}
We have shown that introducing a poly-Si top gate can effectively reduce the level of the charge offset drift in Si/SiO$_2$-based single gate layer SEDs. This clearly demonstrates that the level of charge offset drift measured depends on factors other than simply the material systems used as previously thought, and the device design plays an important role. Not only do these results provide researchers the opportunity to tune the level of stability performance in their devices, it provides an avenue toward further understanding the origin of noise in devices in various material systems.

\section*{SUPPLEMENTARY MATERIAL}
See supplementary material for the equivalent circuit model to deduce the relationship $Q_0(t)\approx (C_m/C_d)\Delta Q_d(t)$, and the FastCap simulation. It also includes addition data not presented in the main text.

\section*{ACKNOWLEDGMENTS}
We are grateful to acknowledge useful discussions with Joshua M. Pomeroy, Joseph A. Hagmann, and Ryan Stein. We would also like to thank Ron Manginell, Malcom Carroll, and co-workers at Sandia National Laboratories for providing us the single layer SED lithographic design. Research was performed in part at the NIST Center for Nanoscale Science and Technology.

\section*{Disclaimer}
Certain commercial equipment, instruments and materials are identified in order to specify experimental procedures as completely as possible. In no case does such identification imply a recommendation or it imply that any of the materials, instruments or equipment identified are necessarily the best available for the purpose.


\begin{thebibliography}{31}%
\makeatletter
\providecommand \@ifxundefined [1]{%
 \@ifx{#1\undefined}
}%
\providecommand \@ifnum [1]{%
 \ifnum #1\expandafter \@firstoftwo
 \else \expandafter \@secondoftwo
 \fi
}%
\providecommand \@ifx [1]{%
 \ifx #1\expandafter \@firstoftwo
 \else \expandafter \@secondoftwo
 \fi
}%
\providecommand \natexlab [1]{#1}%
\providecommand \enquote  [1]{``#1''}%
\providecommand \bibnamefont  [1]{#1}%
\providecommand \bibfnamefont [1]{#1}%
\providecommand \citenamefont [1]{#1}%
\providecommand \href@noop [0]{\@secondoftwo}%
\providecommand \href [0]{\begingroup \@sanitize@url \@href}%
\providecommand \@href[1]{\@@startlink{#1}\@@href}%
\providecommand \@@href[1]{\endgroup#1\@@endlink}%
\providecommand \@sanitize@url [0]{\catcode `\\12\catcode `\$12\catcode
  `\&12\catcode `\#12\catcode `\^12\catcode `\_12\catcode `\%12\relax}%
\providecommand \@@startlink[1]{}%
\providecommand \@@endlink[0]{}%
\providecommand \url  [0]{\begingroup\@sanitize@url \@url }%
\providecommand \@url [1]{\endgroup\@href {#1}{\urlprefix }}%
\providecommand \urlprefix  [0]{URL }%
\providecommand \Eprint [0]{\href }%
\providecommand \doibase [0]{http://dx.doi.org/}%
\providecommand \selectlanguage [0]{\@gobble}%
\providecommand \bibinfo  [0]{\@secondoftwo}%
\providecommand \bibfield  [0]{\@secondoftwo}%
\providecommand \translation [1]{[#1]}%
\providecommand \BibitemOpen [0]{}%
\providecommand \bibitemStop [0]{}%
\providecommand \bibitemNoStop [0]{.\EOS\space}%
\providecommand \EOS [0]{\spacefactor3000\relax}%
\providecommand \BibitemShut  [1]{\csname bibitem#1\endcsname}%
\let\auto@bib@innerbib\@empty
\bibitem [{\citenamefont {Keller}\ \emph {et~al.}(1996)\citenamefont {Keller},
  \citenamefont {Martinis}, \citenamefont {Zimmerman},\ and\ \citenamefont
  {Steinbach}}]{Keller96}%
  \BibitemOpen
  \bibfield  {author} {\bibinfo {author} {\bibfnamefont {M.~W.}\ \bibnamefont
  {Keller}}, \bibinfo {author} {\bibfnamefont {J.~M.}\ \bibnamefont
  {Martinis}}, \bibinfo {author} {\bibfnamefont {N.~M.}\ \bibnamefont
  {Zimmerman}}, \ and\ \bibinfo {author} {\bibfnamefont {A.~H.}\ \bibnamefont
  {Steinbach}},\ }\href {\doibase 10.1063/1.117492} {\bibfield  {journal}
  {\bibinfo  {journal} {Applied Physics Letters}\ }\textbf {\bibinfo {volume}
  {69}},\ \bibinfo {pages} {1804} (\bibinfo {year} {1996})}\BibitemShut
  {NoStop}%
\bibitem [{\citenamefont {Wright}\ \emph {et~al.}(2009)\citenamefont {Wright},
  \citenamefont {Blumenthal}, \citenamefont {Pepper}, \citenamefont {Anderson},
  \citenamefont {Jones}, \citenamefont {Nicoll},\ and\ \citenamefont
  {Ritchie}}]{Wright09}%
  \BibitemOpen
  \bibfield  {author} {\bibinfo {author} {\bibfnamefont {S.~J.}\ \bibnamefont
  {Wright}}, \bibinfo {author} {\bibfnamefont {M.~D.}\ \bibnamefont
  {Blumenthal}}, \bibinfo {author} {\bibfnamefont {M.}~\bibnamefont {Pepper}},
  \bibinfo {author} {\bibfnamefont {D.}~\bibnamefont {Anderson}}, \bibinfo
  {author} {\bibfnamefont {G.~A.~C.}\ \bibnamefont {Jones}}, \bibinfo {author}
  {\bibfnamefont {C.~A.}\ \bibnamefont {Nicoll}}, \ and\ \bibinfo {author}
  {\bibfnamefont {D.~A.}\ \bibnamefont {Ritchie}},\ }\href {\doibase
  10.1103/PhysRevB.80.113303} {\bibfield  {journal} {\bibinfo  {journal} {Phys.
  Rev. B}\ }\textbf {\bibinfo {volume} {80}},\ \bibinfo {pages} {113303}
  (\bibinfo {year} {2009})}\BibitemShut {NoStop}%
\bibitem [{\citenamefont {Maisi}\ \emph {et~al.}(2009)\citenamefont {Maisi},
  \citenamefont {Pashkin}, \citenamefont {Kafanov}, \citenamefont {Tsai},\ and\
  \citenamefont {Pekola}}]{Maisi09}%
  \BibitemOpen
  \bibfield  {author} {\bibinfo {author} {\bibfnamefont {V.~F.}\ \bibnamefont
  {Maisi}}, \bibinfo {author} {\bibfnamefont {Y.~A.}\ \bibnamefont {Pashkin}},
  \bibinfo {author} {\bibfnamefont {S.}~\bibnamefont {Kafanov}}, \bibinfo
  {author} {\bibfnamefont {J.-S.}\ \bibnamefont {Tsai}}, \ and\ \bibinfo
  {author} {\bibfnamefont {J.~P.}\ \bibnamefont {Pekola}},\ }\href
  {http://stacks.iop.org/1367-2630/11/i=11/a=113057} {\bibfield  {journal}
  {\bibinfo  {journal} {New Journal of Physics}\ }\textbf {\bibinfo {volume}
  {11}},\ \bibinfo {pages} {113057} (\bibinfo {year} {2009})}\BibitemShut
  {NoStop}%
\bibitem [{\citenamefont {Fletcher}\ \emph {et~al.}(2012)\citenamefont
  {Fletcher}, \citenamefont {Kataoka}, \citenamefont {Giblin}, \citenamefont
  {Park}, \citenamefont {Sim}, \citenamefont {See}, \citenamefont {Ritchie},
  \citenamefont {Griffiths}, \citenamefont {Jones}, \citenamefont {Beere},\
  and\ \citenamefont {Janssen}}]{Fletcher12}%
  \BibitemOpen
  \bibfield  {author} {\bibinfo {author} {\bibfnamefont {J.~D.}\ \bibnamefont
  {Fletcher}}, \bibinfo {author} {\bibfnamefont {M.}~\bibnamefont {Kataoka}},
  \bibinfo {author} {\bibfnamefont {S.~P.}\ \bibnamefont {Giblin}}, \bibinfo
  {author} {\bibfnamefont {S.}~\bibnamefont {Park}}, \bibinfo {author}
  {\bibfnamefont {H.-S.}\ \bibnamefont {Sim}}, \bibinfo {author} {\bibfnamefont
  {P.}~\bibnamefont {See}}, \bibinfo {author} {\bibfnamefont {D.~A.}\
  \bibnamefont {Ritchie}}, \bibinfo {author} {\bibfnamefont {J.~P.}\
  \bibnamefont {Griffiths}}, \bibinfo {author} {\bibfnamefont {G.~A.~C.}\
  \bibnamefont {Jones}}, \bibinfo {author} {\bibfnamefont {H.~E.}\ \bibnamefont
  {Beere}}, \ and\ \bibinfo {author} {\bibfnamefont {T.~J. B.~M.}\ \bibnamefont
  {Janssen}},\ }\href {\doibase 10.1103/PhysRevB.86.155311} {\bibfield
  {journal} {\bibinfo  {journal} {Phys. Rev. B}\ }\textbf {\bibinfo {volume}
  {86}},\ \bibinfo {pages} {155311} (\bibinfo {year} {2012})}\BibitemShut
  {NoStop}%
\bibitem [{\citenamefont {Giblin}\ \emph {et~al.}(2012)\citenamefont {Giblin},
  \citenamefont {Kataoka}, \citenamefont {Fletcher}, \citenamefont {See},
  \citenamefont {Janssen}, \citenamefont {Griffiths}, \citenamefont {Jones},
  \citenamefont {Farrer},\ and\ \citenamefont {Ritchie}}]{Giblin2012}%
  \BibitemOpen
  \bibfield  {author} {\bibinfo {author} {\bibfnamefont {S.~P.}\ \bibnamefont
  {Giblin}}, \bibinfo {author} {\bibfnamefont {M.}~\bibnamefont {Kataoka}},
  \bibinfo {author} {\bibfnamefont {J.~D.}\ \bibnamefont {Fletcher}}, \bibinfo
  {author} {\bibfnamefont {P.}~\bibnamefont {See}}, \bibinfo {author}
  {\bibfnamefont {T.~J. B.~M.}\ \bibnamefont {Janssen}}, \bibinfo {author}
  {\bibfnamefont {J.~P.}\ \bibnamefont {Griffiths}}, \bibinfo {author}
  {\bibfnamefont {G.~A.~C.}\ \bibnamefont {Jones}}, \bibinfo {author}
  {\bibfnamefont {I.}~\bibnamefont {Farrer}}, \ and\ \bibinfo {author}
  {\bibfnamefont {D.~A.}\ \bibnamefont {Ritchie}},\ }\href
  {http://dx.doi.org/10.1038/ncomms1935} {\bibfield  {journal} {\bibinfo
  {journal} {Nat. Commun.}\ }\textbf {\bibinfo {volume} {3}},\ \bibinfo {pages}
  {930} (\bibinfo {year} {2012})}\BibitemShut {NoStop}%
\bibitem [{\citenamefont {Pekola}\ \emph {et~al.}(2013)\citenamefont {Pekola},
  \citenamefont {Saira}, \citenamefont {Maisi}, \citenamefont {Kemppinen},
  \citenamefont {M\"ott\"onen}, \citenamefont {Pashkin},\ and\ \citenamefont
  {Averin}}]{Pekola13}%
  \BibitemOpen
  \bibfield  {author} {\bibinfo {author} {\bibfnamefont {J.~P.}\ \bibnamefont
  {Pekola}}, \bibinfo {author} {\bibfnamefont {O.-P.}\ \bibnamefont {Saira}},
  \bibinfo {author} {\bibfnamefont {V.~F.}\ \bibnamefont {Maisi}}, \bibinfo
  {author} {\bibfnamefont {A.}~\bibnamefont {Kemppinen}}, \bibinfo {author}
  {\bibfnamefont {M.}~\bibnamefont {M\"ott\"onen}}, \bibinfo {author}
  {\bibfnamefont {Y.~A.}\ \bibnamefont {Pashkin}}, \ and\ \bibinfo {author}
  {\bibfnamefont {D.~V.}\ \bibnamefont {Averin}},\ }\href {\doibase
  10.1103/RevModPhys.85.1421} {\bibfield  {journal} {\bibinfo  {journal} {Rev.
  Mod. Phys.}\ }\textbf {\bibinfo {volume} {85}},\ \bibinfo {pages} {1421}
  (\bibinfo {year} {2013})}\BibitemShut {NoStop}%
\bibitem [{\citenamefont {Yamahata}, \citenamefont {Nishiguchi},\ and\
  \citenamefont {Fujiwara}(2014)}]{Yamahata2014}%
  \BibitemOpen
  \bibfield  {author} {\bibinfo {author} {\bibfnamefont {G.}~\bibnamefont
  {Yamahata}}, \bibinfo {author} {\bibfnamefont {K.}~\bibnamefont
  {Nishiguchi}}, \ and\ \bibinfo {author} {\bibfnamefont {A.}~\bibnamefont
  {Fujiwara}},\ }\href {http://dx.doi.org/10.1038/ncomms6038} {\bibfield
  {journal} {\bibinfo  {journal} {Nat. Commun.}\ }\textbf {\bibinfo {volume}
  {5}},\ \bibinfo {pages} {5038} (\bibinfo {year} {2014})}\BibitemShut
  {NoStop}%
\bibitem [{\citenamefont {Nakamura}, \citenamefont {Tsai},\ and\ \citenamefont
  {Kaneko}(2015)}]{Nakamura15}%
  \BibitemOpen
  \bibfield  {author} {\bibinfo {author} {\bibfnamefont {Y.}~\bibnamefont
  {Nakamura}, \bibfnamefont {S.and~Pashkin}}, \bibinfo {author} {\bibfnamefont
  {J.}~\bibnamefont {Tsai}}, \ and\ \bibinfo {author} {\bibfnamefont
  {N.}~\bibnamefont {Kaneko}},\ }\href@noop {} {\bibfield  {journal} {\bibinfo
  {journal} {IEEE Trans. Instrum. Meas.}\ }\textbf {\bibinfo {volume} {64}},\
  \bibinfo {pages} {1696–1701} (\bibinfo {year} {2015})}\BibitemShut
  {NoStop}%
\bibitem [{\citenamefont {Yamahata}, \citenamefont {Karasawa},\ and\
  \citenamefont {Fujiwara}(2015)}]{Yamahata15}%
  \BibitemOpen
  \bibfield  {author} {\bibinfo {author} {\bibfnamefont {G.}~\bibnamefont
  {Yamahata}}, \bibinfo {author} {\bibfnamefont {T.}~\bibnamefont {Karasawa}},
  \ and\ \bibinfo {author} {\bibfnamefont {A.}~\bibnamefont {Fujiwara}},\
  }\href {\doibase 10.1063/1.4905934} {\bibfield  {journal} {\bibinfo
  {journal} {Applied Physics Letters}\ }\textbf {\bibinfo {volume} {106}},\
  \bibinfo {pages} {023112} (\bibinfo {year} {2015})}\BibitemShut {NoStop}%
\bibitem [{\citenamefont {Chen}, \citenamefont {Korotkov},\ and\ \citenamefont
  {Likharev}(1995)}]{chen95}%
  \BibitemOpen
  \bibfield  {author} {\bibinfo {author} {\bibfnamefont {R.~H.}\ \bibnamefont
  {Chen}}, \bibinfo {author} {\bibfnamefont {A.~N.}\ \bibnamefont {Korotkov}},
  \ and\ \bibinfo {author} {\bibfnamefont {K.~K.}\ \bibnamefont {Likharev}},\
  }in\ \href {\doibase 10.1109/DRC.1995.496242} {\emph {\bibinfo {booktitle}
  {1995 53rd Annual Device Research Conference Digest}}}\ (\bibinfo {year}
  {1995})\ pp.\ \bibinfo {pages} {44--45}\BibitemShut {NoStop}%
\bibitem [{\citenamefont {Likharev}(1999)}]{Likharev99}%
  \BibitemOpen
  \bibfield  {author} {\bibinfo {author} {\bibfnamefont {K.~K.}\ \bibnamefont
  {Likharev}},\ }\href {\doibase 10.1109/5.752518} {\bibfield  {journal}
  {\bibinfo  {journal} {Proceedings of the IEEE}\ }\textbf {\bibinfo {volume}
  {87}},\ \bibinfo {pages} {606} (\bibinfo {year} {1999})}\BibitemShut
  {NoStop}%
\bibitem [{\citenamefont {Takahashi}\ \emph {et~al.}(2002)\citenamefont
  {Takahashi}, \citenamefont {Ono}, \citenamefont {Fujiwara},\ and\
  \citenamefont {Inokawa}}]{Takahashi02}%
  \BibitemOpen
  \bibfield  {author} {\bibinfo {author} {\bibfnamefont {Y.}~\bibnamefont
  {Takahashi}}, \bibinfo {author} {\bibfnamefont {Y.}~\bibnamefont {Ono}},
  \bibinfo {author} {\bibfnamefont {A.}~\bibnamefont {Fujiwara}}, \ and\
  \bibinfo {author} {\bibfnamefont {H.}~\bibnamefont {Inokawa}},\ }\href
  {http://stacks.iop.org/0953-8984/14/i=39/a=201} {\bibfield  {journal}
  {\bibinfo  {journal} {Journal of Physics: Condensed Matter}\ }\textbf
  {\bibinfo {volume} {14}},\ \bibinfo {pages} {R995} (\bibinfo {year}
  {2002})}\BibitemShut {NoStop}%
\bibitem [{\citenamefont {Takahashi}\ \emph {et~al.}(2004)\citenamefont
  {Takahashi}, \citenamefont {Ono}, \citenamefont {Fujiwara},\ and\
  \citenamefont {Inokawa}}]{Takahashi04}%
  \BibitemOpen
  \bibfield  {author} {\bibinfo {author} {\bibfnamefont {Y.}~\bibnamefont
  {Takahashi}}, \bibinfo {author} {\bibfnamefont {Y.}~\bibnamefont {Ono}},
  \bibinfo {author} {\bibfnamefont {A.}~\bibnamefont {Fujiwara}}, \ and\
  \bibinfo {author} {\bibfnamefont {H.}~\bibnamefont {Inokawa}},\ }in\ \href
  {\doibase 10.1109/ICSICT.2004.1435083} {\emph {\bibinfo {booktitle}
  {Proceedings. 7th International Conference on Solid-State and Integrated
  Circuits Technology, 2004.}}},\ Vol.~\bibinfo {volume} {1}\ (\bibinfo {year}
  {2004})\ pp.\ \bibinfo {pages} {624--629 vol.1}\BibitemShut {NoStop}%
\bibitem [{\citenamefont {Maeda}\ \emph {et~al.}(2012)\citenamefont {Maeda},
  \citenamefont {Okabayashi}, \citenamefont {Kano}, \citenamefont {Takeshita},
  \citenamefont {Tanaka}, \citenamefont {Sakamoto}, \citenamefont {Teranishi},\
  and\ \citenamefont {Majima}}]{Maeda12}%
  \BibitemOpen
  \bibfield  {author} {\bibinfo {author} {\bibfnamefont {K.}~\bibnamefont
  {Maeda}}, \bibinfo {author} {\bibfnamefont {N.}~\bibnamefont {Okabayashi}},
  \bibinfo {author} {\bibfnamefont {S.}~\bibnamefont {Kano}}, \bibinfo {author}
  {\bibfnamefont {S.}~\bibnamefont {Takeshita}}, \bibinfo {author}
  {\bibfnamefont {D.}~\bibnamefont {Tanaka}}, \bibinfo {author} {\bibfnamefont
  {M.}~\bibnamefont {Sakamoto}}, \bibinfo {author} {\bibfnamefont
  {T.}~\bibnamefont {Teranishi}}, \ and\ \bibinfo {author} {\bibfnamefont
  {Y.}~\bibnamefont {Majima}},\ }\href {\doibase 10.1021/nn3003086} {\bibfield
  {journal} {\bibinfo  {journal} {ACS Nano}\ }\textbf {\bibinfo {volume} {6}},\
  \bibinfo {pages} {2798} (\bibinfo {year} {2012})}\BibitemShut {NoStop}%
\bibitem [{\citenamefont {Loss}\ and\ \citenamefont
  {DiVincenzo}(1998)}]{Loss98}%
  \BibitemOpen
  \bibfield  {author} {\bibinfo {author} {\bibfnamefont {D.}~\bibnamefont
  {Loss}}\ and\ \bibinfo {author} {\bibfnamefont {D.~P.}\ \bibnamefont
  {DiVincenzo}},\ }\href {\doibase 10.1103/PhysRevA.57.120} {\bibfield
  {journal} {\bibinfo  {journal} {Phys. Rev. A}\ }\textbf {\bibinfo {volume}
  {57}},\ \bibinfo {pages} {120} (\bibinfo {year} {1998})}\BibitemShut
  {NoStop}%
\bibitem [{\citenamefont {Kane}(1998)}]{Kane1998}%
  \BibitemOpen
  \bibfield  {author} {\bibinfo {author} {\bibfnamefont {B.~E.}\ \bibnamefont
  {Kane}},\ }\href {\doibase 10.1038/30156} {\bibfield  {journal} {\bibinfo
  {journal} {Nature}\ }\textbf {\bibinfo {volume} {393}},\ \bibinfo {pages}
  {133} (\bibinfo {year} {1998})}\BibitemShut {NoStop}%
\bibitem [{\citenamefont {van~der Wiel}\ \emph {et~al.}(2002)\citenamefont
  {van~der Wiel}, \citenamefont {De~Franceschi}, \citenamefont {Elzerman},
  \citenamefont {Fujisawa}, \citenamefont {Tarucha},\ and\ \citenamefont
  {Kouwenhoven}}]{Wiel02}%
  \BibitemOpen
  \bibfield  {author} {\bibinfo {author} {\bibfnamefont {W.~G.}\ \bibnamefont
  {van~der Wiel}}, \bibinfo {author} {\bibfnamefont {S.}~\bibnamefont
  {De~Franceschi}}, \bibinfo {author} {\bibfnamefont {J.~M.}\ \bibnamefont
  {Elzerman}}, \bibinfo {author} {\bibfnamefont {T.}~\bibnamefont {Fujisawa}},
  \bibinfo {author} {\bibfnamefont {S.}~\bibnamefont {Tarucha}}, \ and\
  \bibinfo {author} {\bibfnamefont {L.~P.}\ \bibnamefont {Kouwenhoven}},\
  }\href {\doibase 10.1103/RevModPhys.75.1} {\bibfield  {journal} {\bibinfo
  {journal} {Rev. Mod. Phys.}\ }\textbf {\bibinfo {volume} {75}},\ \bibinfo
  {pages} {1} (\bibinfo {year} {2002})}\BibitemShut {NoStop}%
\bibitem [{\citenamefont {Hanson}\ \emph {et~al.}(2007)\citenamefont {Hanson},
  \citenamefont {Kouwenhoven}, \citenamefont {Petta}, \citenamefont {Tarucha},\
  and\ \citenamefont {Vandersypen}}]{Hanson07}%
  \BibitemOpen
  \bibfield  {author} {\bibinfo {author} {\bibfnamefont {R.}~\bibnamefont
  {Hanson}}, \bibinfo {author} {\bibfnamefont {L.~P.}\ \bibnamefont
  {Kouwenhoven}}, \bibinfo {author} {\bibfnamefont {J.~R.}\ \bibnamefont
  {Petta}}, \bibinfo {author} {\bibfnamefont {S.}~\bibnamefont {Tarucha}}, \
  and\ \bibinfo {author} {\bibfnamefont {L.~M.~K.}\ \bibnamefont
  {Vandersypen}},\ }\href {\doibase 10.1103/RevModPhys.79.1217} {\bibfield
  {journal} {\bibinfo  {journal} {Rev. Mod. Phys.}\ }\textbf {\bibinfo {volume}
  {79}},\ \bibinfo {pages} {1217} (\bibinfo {year} {2007})}\BibitemShut
  {NoStop}%
\bibitem [{\citenamefont {Zwanenburg}\ \emph {et~al.}(2013)\citenamefont
  {Zwanenburg}, \citenamefont {Dzurak}, \citenamefont {Morello}, \citenamefont
  {Simmons}, \citenamefont {Hollenberg}, \citenamefont {Klimeck}, \citenamefont
  {Rogge}, \citenamefont {Coppersmith},\ and\ \citenamefont
  {Eriksson}}]{Zwanenburg13}%
  \BibitemOpen
  \bibfield  {author} {\bibinfo {author} {\bibfnamefont {F.~A.}\ \bibnamefont
  {Zwanenburg}}, \bibinfo {author} {\bibfnamefont {A.~S.}\ \bibnamefont
  {Dzurak}}, \bibinfo {author} {\bibfnamefont {A.}~\bibnamefont {Morello}},
  \bibinfo {author} {\bibfnamefont {M.~Y.}\ \bibnamefont {Simmons}}, \bibinfo
  {author} {\bibfnamefont {L.~C.~L.}\ \bibnamefont {Hollenberg}}, \bibinfo
  {author} {\bibfnamefont {G.}~\bibnamefont {Klimeck}}, \bibinfo {author}
  {\bibfnamefont {S.}~\bibnamefont {Rogge}}, \bibinfo {author} {\bibfnamefont
  {S.~N.}\ \bibnamefont {Coppersmith}}, \ and\ \bibinfo {author} {\bibfnamefont
  {M.~A.}\ \bibnamefont {Eriksson}},\ }\href {\doibase
  10.1103/RevModPhys.85.961} {\bibfield  {journal} {\bibinfo  {journal} {Rev.
  Mod. Phys.}\ }\textbf {\bibinfo {volume} {85}},\ \bibinfo {pages} {961}
  (\bibinfo {year} {2013})}\BibitemShut {NoStop}%
\bibitem [{\citenamefont {Stewart}\ and\ \citenamefont
  {Zimmerman}(2016)}]{Stew16}%
  \BibitemOpen
  \bibfield  {author} {\bibinfo {author} {\bibfnamefont {M.~D.}\ \bibnamefont
  {Stewart}}\ and\ \bibinfo {author} {\bibfnamefont {N.~M.}\ \bibnamefont
  {Zimmerman}},\ }\href {\doibase 10.3390/app6070187} {\bibfield  {journal}
  {\bibinfo  {journal} {Applied Sciences}\ }\textbf {\bibinfo {volume} {6}},\
  \bibinfo {pages} {187} (\bibinfo {year} {2016})}\BibitemShut {NoStop}%
\bibitem [{\citenamefont {Zimmerman}\ \emph {et~al.}(2008)\citenamefont
  {Zimmerman}, \citenamefont {Huber}, \citenamefont {Simonds}, \citenamefont
  {Hourdakis}, \citenamefont {Fujiwara}, \citenamefont {Ono}, \citenamefont
  {Takahashi}, \citenamefont {Inokawa}, \citenamefont {Furlan},\ and\
  \citenamefont {Keller}}]{Neil08JAP}%
  \BibitemOpen
  \bibfield  {author} {\bibinfo {author} {\bibfnamefont {N.~M.}\ \bibnamefont
  {Zimmerman}}, \bibinfo {author} {\bibfnamefont {W.~H.}\ \bibnamefont
  {Huber}}, \bibinfo {author} {\bibfnamefont {B.}~\bibnamefont {Simonds}},
  \bibinfo {author} {\bibfnamefont {E.}~\bibnamefont {Hourdakis}}, \bibinfo
  {author} {\bibfnamefont {A.}~\bibnamefont {Fujiwara}}, \bibinfo {author}
  {\bibfnamefont {Y.}~\bibnamefont {Ono}}, \bibinfo {author} {\bibfnamefont
  {Y.}~\bibnamefont {Takahashi}}, \bibinfo {author} {\bibfnamefont
  {H.}~\bibnamefont {Inokawa}}, \bibinfo {author} {\bibfnamefont
  {M.}~\bibnamefont {Furlan}}, \ and\ \bibinfo {author} {\bibfnamefont {M.~W.}\
  \bibnamefont {Keller}},\ }\href {\doibase 10.1063/1.2949700} {\bibfield
  {journal} {\bibinfo  {journal} {Journal of Applied Physics}\ }\textbf
  {\bibinfo {volume} {104}},\ \bibinfo {pages} {033710} (\bibinfo {year}
  {2008})}\BibitemShut {NoStop}%
\bibitem [{\citenamefont {Huber}, \citenamefont {Martin},\ and\ \citenamefont
  {Zimmerman}(2001)}]{Huber01}%
  \BibitemOpen
  \bibfield  {author} {\bibinfo {author} {\bibfnamefont {W.~H.}\ \bibnamefont
  {Huber}}, \bibinfo {author} {\bibfnamefont {S.~B.}\ \bibnamefont {Martin}}, \
  and\ \bibinfo {author} {\bibfnamefont {N.~M.}\ \bibnamefont {Zimmerman}},\
  }in\ \href@noop {} {\emph {\bibinfo {booktitle} {Experimental Implementation
  of Quantum Computation (IQC’01)}}},\ Vol.~\bibinfo {volume} {76},\ \bibinfo
  {editor} {edited by\ \bibinfo {editor} {\bibfnamefont {R.}~\bibnamefont
  {Clark}}}\ (\bibinfo {address} {Rinton Press, NJ},\ \bibinfo {year}
  {2001})\BibitemShut {NoStop}%
\bibitem [{\citenamefont {Zimmerman}\ \emph {et~al.}(2001)\citenamefont
  {Zimmerman}, \citenamefont {Huber}, \citenamefont {Fujiwara},\ and\
  \citenamefont {Takahashi}}]{Neil2001}%
  \BibitemOpen
  \bibfield  {author} {\bibinfo {author} {\bibfnamefont {N.~M.}\ \bibnamefont
  {Zimmerman}}, \bibinfo {author} {\bibfnamefont {W.~H.}\ \bibnamefont
  {Huber}}, \bibinfo {author} {\bibfnamefont {A.}~\bibnamefont {Fujiwara}}, \
  and\ \bibinfo {author} {\bibfnamefont {Y.}~\bibnamefont {Takahashi}},\ }\href
  {\doibase 10.1063/1.1415776} {\bibfield  {journal} {\bibinfo  {journal}
  {Applied Physics Letters}\ }\textbf {\bibinfo {volume} {79}},\ \bibinfo
  {pages} {3188} (\bibinfo {year} {2001})}\BibitemShut {NoStop}%
\bibitem [{\citenamefont {Zimmerman}\ \emph {et~al.}(2007)\citenamefont
  {Zimmerman}, \citenamefont {Simonds}, \citenamefont {Fujiwara}, \citenamefont
  {Ono}, \citenamefont {Takahashi},\ and\ \citenamefont {Inokawa}}]{Neil2007}%
  \BibitemOpen
  \bibfield  {author} {\bibinfo {author} {\bibfnamefont {N.~M.}\ \bibnamefont
  {Zimmerman}}, \bibinfo {author} {\bibfnamefont {B.~J.}\ \bibnamefont
  {Simonds}}, \bibinfo {author} {\bibfnamefont {A.}~\bibnamefont {Fujiwara}},
  \bibinfo {author} {\bibfnamefont {Y.}~\bibnamefont {Ono}}, \bibinfo {author}
  {\bibfnamefont {Y.}~\bibnamefont {Takahashi}}, \ and\ \bibinfo {author}
  {\bibfnamefont {H.}~\bibnamefont {Inokawa}},\ }\href {\doibase
  10.1063/1.2431778} {\bibfield  {journal} {\bibinfo  {journal} {Applied
  Physics Letters}\ }\textbf {\bibinfo {volume} {90}},\ \bibinfo {pages}
  {033507} (\bibinfo {year} {2007})}\BibitemShut {NoStop}%
\bibitem [{\citenamefont {Zimmerman}\ \emph {et~al.}(2014)\citenamefont
  {Zimmerman}, \citenamefont {Yang}, \citenamefont {Lai}, \citenamefont {Lim},\
  and\ \citenamefont {Dzurak}}]{Neil2014}%
  \BibitemOpen
  \bibfield  {author} {\bibinfo {author} {\bibfnamefont {N.~M.}\ \bibnamefont
  {Zimmerman}}, \bibinfo {author} {\bibfnamefont {C.-H.}\ \bibnamefont {Yang}},
  \bibinfo {author} {\bibfnamefont {N.~S.}\ \bibnamefont {Lai}}, \bibinfo
  {author} {\bibfnamefont {W.~H.}\ \bibnamefont {Lim}}, \ and\ \bibinfo
  {author} {\bibfnamefont {A.~S.}\ \bibnamefont {Dzurak}},\ }\href
  {http://stacks.iop.org/0957-4484/25/i=40/a=405201} {\bibfield  {journal}
  {\bibinfo  {journal} {Nanotechnology}\ }\textbf {\bibinfo {volume} {25}},\
  \bibinfo {pages} {405201} (\bibinfo {year} {2014})}\BibitemShut {NoStop}%
\bibitem [{\citenamefont {Freeman}, \citenamefont {Schoenfield},\ and\
  \citenamefont {Jiang}(2016)}]{Blake16}%
  \BibitemOpen
  \bibfield  {author} {\bibinfo {author} {\bibfnamefont {B.~M.}\ \bibnamefont
  {Freeman}}, \bibinfo {author} {\bibfnamefont {J.~S.}\ \bibnamefont
  {Schoenfield}}, \ and\ \bibinfo {author} {\bibfnamefont {H.}~\bibnamefont
  {Jiang}},\ }\href {\doibase 10.1063/1.4954700} {\bibfield  {journal}
  {\bibinfo  {journal} {Applied Physics Letters}\ }\textbf {\bibinfo {volume}
  {108}},\ \bibinfo {pages} {253108} (\bibinfo {year} {2016})},\ \Eprint
  {http://arxiv.org/abs/https://doi.org/10.1063/1.4954700}
  {https://doi.org/10.1063/1.4954700} \BibitemShut {NoStop}%
\bibitem [{\citenamefont {Tracy}\ \emph {et~al.}(2013)\citenamefont {Tracy},
  \citenamefont {Lu}, \citenamefont {Bishop}, \citenamefont {Eyck},
  \citenamefont {Pluym}, \citenamefont {Wendt}, \citenamefont {Lilly},\ and\
  \citenamefont {Carroll}}]{Tracy13}%
  \BibitemOpen
  \bibfield  {author} {\bibinfo {author} {\bibfnamefont {L.~A.}\ \bibnamefont
  {Tracy}}, \bibinfo {author} {\bibfnamefont {T.~M.}\ \bibnamefont {Lu}},
  \bibinfo {author} {\bibfnamefont {N.~C.}\ \bibnamefont {Bishop}}, \bibinfo
  {author} {\bibfnamefont {G.~A.~T.}\ \bibnamefont {Eyck}}, \bibinfo {author}
  {\bibfnamefont {T.}~\bibnamefont {Pluym}}, \bibinfo {author} {\bibfnamefont
  {J.~R.}\ \bibnamefont {Wendt}}, \bibinfo {author} {\bibfnamefont {M.~P.}\
  \bibnamefont {Lilly}}, \ and\ \bibinfo {author} {\bibfnamefont {M.~S.}\
  \bibnamefont {Carroll}},\ }\href {\doibase 10.1063/1.4824128} {\bibfield
  {journal} {\bibinfo  {journal} {Applied Physics Letters}\ }\textbf {\bibinfo
  {volume} {103}},\ \bibinfo {pages} {143115} (\bibinfo {year}
  {2013})}\BibitemShut {NoStop}%
\bibitem [{\citenamefont {Rudolph}\ \emph {et~al.}()\citenamefont {Rudolph},
  \citenamefont {Sarabi}, \citenamefont {Murray}, \citenamefont {Carroll},\
  and\ \citenamefont {Zimmerman}}]{Rudolph17}%
  \BibitemOpen
  \bibfield  {author} {\bibinfo {author} {\bibfnamefont {M.}~\bibnamefont
  {Rudolph}}, \bibinfo {author} {\bibfnamefont {B.}~\bibnamefont {Sarabi}},
  \bibinfo {author} {\bibfnamefont {R.}~\bibnamefont {Murray}}, \bibinfo
  {author} {\bibfnamefont {M.~S.}\ \bibnamefont {Carroll}}, \ and\ \bibinfo
  {author} {\bibfnamefont {N.}~\bibnamefont {Zimmerman}},\ }\href@noop {}
  {\bibinfo  {journal} {arXiv:1801.07776}\ }\BibitemShut {NoStop}%
\bibitem [{\citenamefont {Rossi}\ \emph {et~al.}(2015)\citenamefont {Rossi},
  \citenamefont {Tanttu}, \citenamefont {Hudson}, \citenamefont {Sun},
  \citenamefont {M{\"o}tt{\"o}nen},\ and\ \citenamefont {Dzurak}}]{Rossi2015}%
  \BibitemOpen
\bibfield  {journal} {  }\bibfield  {author} {\bibinfo {author} {\bibfnamefont
  {A.}~\bibnamefont {Rossi}}, \bibinfo {author} {\bibfnamefont
  {T.}~\bibnamefont {Tanttu}}, \bibinfo {author} {\bibfnamefont {F.~E.}\
  \bibnamefont {Hudson}}, \bibinfo {author} {\bibfnamefont {Y.}~\bibnamefont
  {Sun}}, \bibinfo {author} {\bibfnamefont {M.}~\bibnamefont
  {M{\"o}tt{\"o}nen}}, \ and\ \bibinfo {author} {\bibfnamefont {A.~S.}\
  \bibnamefont {Dzurak}},\ }\href {\doibase 10.3791/52852} {\bibfield
  {journal} {\bibinfo  {journal} {J. Vis. Exp.}\ ,\ \bibinfo {pages} {52852}}
  (\bibinfo {year} {2015})}\BibitemShut {NoStop}%
\bibitem [{\citenamefont {Hu}\ and\ \citenamefont {Yang}(2009)}]{binhui09}%
  \BibitemOpen
  \bibfield  {author} {\bibinfo {author} {\bibfnamefont {B.}~\bibnamefont
  {Hu}}\ and\ \bibinfo {author} {\bibfnamefont {C.~H.}\ \bibnamefont {Yang}},\
  }\href {\doibase 10.1103/PhysRevB.80.075310} {\bibfield  {journal} {\bibinfo
  {journal} {Phys. Rev. B}\ }\textbf {\bibinfo {volume} {80}},\ \bibinfo
  {pages} {075310} (\bibinfo {year} {2009})}\BibitemShut {NoStop}%
\bibitem [{\citenamefont {Nabors}\ and\ \citenamefont
  {White}(1991)}]{Nabors91}%
  \BibitemOpen
  \bibfield  {author} {\bibinfo {author} {\bibfnamefont {K.}~\bibnamefont
  {Nabors}}\ and\ \bibinfo {author} {\bibfnamefont {J.}~\bibnamefont {White}},\
  }\href {\doibase 10.1109/43.97624} {\bibfield  {journal} {\bibinfo  {journal}
  {IEEE Trans. Comp.-Aided Des.}\ }\textbf {\bibinfo {volume} {10}},\ \bibinfo
  {pages} {1447} (\bibinfo {year} {1991})}\BibitemShut {NoStop}%
\end{thebibliography}
\end{document}